# Modelling the Structure of a Protein Domain (N-terminal of XPB) Linked with Protein Synthesis, DNA Damage Repair, Rare Diseases, Cancer Therapeutics, and Tuberculosis

**Mitul Saha***

In this work, we develop first near-complete 3D models for NTD-hXPB - the N-terminal protein domain of the human transcription factor XPB. The results are very significant as NTD-hXPB plays a critical role in the synthesis of proteins (specifically transcription) and DNA damage repair (specifically nucleotide excision repair). NTD-hXPB is directly implicated in rare diseases XP-B, XP-CS, and TTD2, whose symptoms include neurodegenerative disorders, premature aging, and decreased fertility. NTD-hXPB is also linked to anti-cancer therapies. As a bi-product we derived 3D models of NTD-mXPB - homologue of NTD-hXPB in mycobacterium tuberculosis *aka* MTB (causative agent of most cases of tuberculosis, which surpassed HIV as #1 infectious disease killer in 2014). These could be potential target for TB therapeutics. Our ab-initio modelling protocol takes advantage of recent powerful advances (prediction of contact residues) in protein structure modeling. Our protocol also includes human in the loop, inspired by the prevailing theory that computational abilities of human minds can be powerfully harnessed in engineering/problem-solving. Using the developed models in this paper, we are able to propose significant insights into (a) the role of NTD-hXPB in the process of transcription and DNA damage repair (specifically NER) (b) the critical interactions of NTD-hXPB with another "co-labourer" protein p52, (c) diseases associated with NTD-hXPB, and (d) alterations in functionalities between NTD-hXPB (in human) and NTD-mXPB (in TB pathogen).

(*) Significant portion of this work done as a postdoctoral fellow at the Sealy Center of Structural Biology, University of Texas Medical Branch at Galveston (http://www.utmb.edu). Contact: misaha@utmb.edu

## INTRODUCTION

Cell is the building block of life and proteins are responsible for almost all tasks in a cell. In order to carry out it's task, a protein must fold into a three dimensional (3D) structure. Hence knowing and understanding the 3D structure of proteins is essential for a complete understanding of life processes, down to the molecular level.

There are established experimental techniques (X-ray, NMR, cryo-EM) to determine the 3D structure of a protein. However they can be quite expensive, time consuming, and also fail at times. As a result only a very small portion of known sequences (over two million) have experimentally determined 3D structures[1]. Hence, computational protein structure prediction methods and approaches can potentially be faster, easier, and affordable alternatives to experimental methods.

In this paper, we use a novel semi-automated computational structure prediction approach to model the 3D structure of NTD-hXPB (N-terminal domain of the human XPB aka xeroderma pigmentosum type B protein – http://www.uniprot.org/uniprot/P19447; residues: 71-200). XPB, and hence NTD-hXPB, are very significant because of number of reasons. XPB plays a critical role in transcription[2]- the first step in synthesis of proteins (recall, proteins perform almost all tasks in the cell). XPB also plays a significant role in DNA damage repair (specifically nucleotide excision repair – NER[3]), which in turn is linked to cancer and aging[4,5]. XPB, along with basal transcription machinery, are also prospective targets of anti-cancer therapies[6,7].

NTD-hXPB has a homologue in *mycobacterium tuberculosis* (MTB)[8]. The structure of the homologue, derived using the model for NTD-hXPB, could be potential therapeutic target for the infectious *tuberculosis (*TB) disease caused by MTB. According to WHO, TB surpassed HIV as #1 infectious disease killer in 2014[9].

NTD-hXPB is implicated in the xeroderma pigmentosum complementation group B (XP-B) - a rare, autosomal recessive disease[10]. The disease leads to reduced DNA repair rates and basal transcription activity. XP-B is known to co-occur with cockayne syndrome (CS) - another rare autosomal recessive. XP-CS patients can experience neurodegenerative disorders characterized by growth failure, impaired development of the nervous system, eye disorders and premature aging. NTD-hXPB is implicated in yet another rare autosomal recessive disorder known as Trichothiodystrophy-2 (TTD2)[11]. TTD patients can display a wide variety of clinical features, including cutaneous, neurologic, growth abnormalities, ichthyosis, intellectual/developmental disabilities, decreased fertility, abnormal characteristics at birth, ocular abnormalities, short stature, and infections.

In this paper we developed near-complete 3D models (TM-Score<0.5) for NTD-hXPB, towards understanding it's role in transcription, DNA repair, and the diseases it is implicated in. We used a novel semi-automated computational approach, based on recent successes with contact-maps, that also involved manual/human intervention/judgements. The approach involving use of human judgements was devised following the failure of established ab-initio fully-automated protein prediction computational methods (ROSETTA, QUARK, EVFOLD) to satisfactorily model NTD-hXPB. Ab-initio approach was needed here as NTD-hXPB has no known structural homologue. The great promises and potential of methodologies involving human intervention/judgement, especially when fully-automated means fail, have been confirmed in other works:- (a) FoldIt approach “to apply the human brain's three-dimensional pattern matching and spatial reasoning abilities to help solve the problem of protein structure prediction” (as per https://en.wikipedia.org/wiki/Foldit, [12]) (b) Luis von Ahn’s works ([13–15], https://www.cs.cmu.edu/~biglou/) in which computational abilities (or "cycles") of humans are used to invent novel techniques, and (c) Murzin’s works with SCOP (http://scop.mrc-lmb.cam.ac.uk/scop/), in which human judgment and expertise is used to analyze protein folds.

To solve the structure of NTD-hXPB we used the computational route after observing that years of experimental research on XPB had only yielded so far the 3D structure of XPB minus NTD-hXPB[8,10,16].

Using the developed models, we propose significant insights into (a) the role of NTD-hXPB in the process of transcription and DNA damage repair (specifically NER) (b) the critical interactions of NTD-hXPB with another protein p52, and (c) rare diseases associated with NTD-hXPB. We also derive and report NTD-mXPB - the structure of the homologue of NTD-hXPB in the tuberculosis pathogen MTB. We compare the two structural homologue pair, NTD-hXPB and NTD-mXPB, especially along the lines of conservation analysis of residue groups in them. We show how the comparison can explain alterations in the functionality between the homologue pair. We also point to potential TB therapeutic target in the NTD-mXPB structure. Finally we also derive and report “an approximate” structure of NTD-hXPB bound with a generic DNA strand.

## METHODOLOGY

This section describes how 3D models for the protein domain NTD-hXPB was derived using semi-automated means. We drew from a number of computational structure modelling/prediction tools :- (a) CHIMERA[17] for visual and secondary structure element (SSE) modelling, (b) MODELLER[18] for loop modelling, (c) JPRED[19], PSIPRED[20], RAPTORX[21] for SSE predictions, (d) BBCONTACTS[22] for beta sheet pairing predictions, (e) NETTURNP[23] for beta turn type predictions, (f) CCMPRED[24], RAPTORX [25] for contact residue pair predictions, and (g) CONSURF[26] for residue conservation statistics.

In this work, we used combined residue-residue contact predictions ("top L/2") from RAPTORX and CCMPRED. For each run, CONSURF performed calculations on 150 sequences with the lowest E-value. It used UNIREF90 and MAFFT for creating alignments. Due to large dataset sizes, we uploaded the residue-residue contact predictions from RAPTORX/CCMPRED and also the CONSURF results at:- http://cs.stanford.edu/~mitul/research/ntd_xpb/, instead of attaching it with Supplementary Information.

### Building 3D models for NTD-hXPB

Supplementary Figure 1 lists secondary structure element predictions for the NTD-hXPB protein sequence (residues 71-200 of protein XPB - http://www.uniprot.org/uniprot/P19447). From the figure, for the NTD-hXPB sequence, predicted secondary structure element sequence SSES is (1E, 2E, 3H, 4E, 5H, 6H, 7H, 8E, 9E, 10H). SSES(i) denotes the i-th element of the sequence. For example, SSES(3) is 3H, and so on.

For the sake of simplicity, we split the NTD-hXPB sequence into three parts (also labeled in the figure):- Motif-1 (M1): residue 71-119, Motif-2 (M2):120-162, and Motif-3 (M3): residues 163-200. The motif partition boundaries were decided after observing that: (a) as per BBCONTACT results, first three predicted sheet elements 1E, 2E, 4E formed a sheet group (details to follow) and (b) 7H formed a strong "coupling" with the previous SSE (6H), which in turn forms helix-turn-helix motif with the SSE before it (details to follow).

We then modelled each of the three motif separately, as described below. After which, we assembled the three motifs against each other to derive two final candidate 3D models for NTD-hXPB. Subsequently, we provide justifications for the finally derived 3D models.

Modelling Motif-1 (residues 71-119)

1. The beta hairpin formed by the two predicted sheets (1E, 2E), as predicted by BBCONTACTS (Supplementary Figure 2), was modelled using known geometrical properties of beta sheets and beta-hairpins. NETTURNP was used to infer the beta-hairpin turn type (Supplementary Figure 3).
2. Then, from SSES, the third SSE (3H; a helix whose 3D helical structure generated by CHIMERA) was manually placed (in CHIMERA visual modeller) in the 3D space against the previous SSE (2E). The placement was done in such a way that most of the predicted contacts (from αC-αC contact predictors, such as, CCMPRED, RAPTORX) were satisfied (αC-αC distance ~8 Å) to the best of the human operator's judgement. This is illustrated in Figure 1 (b).

3. Then the loop connecting this helix and the previous SSE, was modelled using the MODELLER module in CHIMERA.
4. The results of BBCONTACTS was used to "pair" or geometrically model/place the fourth SSE, a sheet element, with the second SSE (another sheet).
5. Then the loop connecting this SSE and the previous SSE, was modelled using MODELLER.
6. The above steps are illustrated in Figure 1.

Modelling Motif-2 (residues 120-162; modeled along the same lines as Motif-1)

1. Having known that this motif contains a structurally conserved DNA binding helix-turn-helix (DB-HtH)[10], we pulled out a DB-HtH structural motif from the protein structures which are known to contain DB-HtH (PDBID: 6CRO, 4CRO, 1LMB, etc). This DB-HtH motif was used as a model for helix-helix structural motif formed by helices 5H (SSES(5)) and 6H (SSES(6)).
2. Then the following helix (SSES(7): 7H) was placed by a human in the 3D space, with respect to previous SSE, in the same way as 3H was placed by a human in a previous step.
3. Again, the loop connecting this SSE to the previous SSE, was modelled using MODELLER.
4. Figure 2 (a) shows Motif-2 with predicted residue pair contacts labeled (via line segments).

Modelling Motif-3 (residues 163-200)

Here we re-used the corresponding structural motif from an attempt of QUARK ab-initio structure prediction of NTD-hXPB, after observing that the residue contact predictions were mostly satisfied in them. Figure 2 (b) shows Motif-3.

Modelling Motif-12

1. Motif-2 was manually placed, in the 3D space, *wrt* Motif-1, in such a way that most of the predicted residue-residue contacts, between them, were satisfied to the best of human operator's judgement. This is done just like manual SSE placements in previous steps.
2. Then, here also, loop connecting Motif-1 and Motif-2 was modelled using MODELLER.
3. We call the resulting motif as Motif-12. Figure 3 illustrates these steps.

Modelling Motif-123 (Final NTD-hXPB structure)

Here we noticed ambiguity. The residue-residue contact predictions between Motif-12 and Motif-3, using CCMPRED, were not confirmed by RAPTORX (hence the predicted residue-residue contacts from CCMPRED could be true or just noise). Nevertheless, we generate a Motif-123 using contacts predicted by CCMPRED (Motif-3 was placed with respect to Motif-12 using the same principles used to place Motif-2 with respect to Motif-1 in the previous "Modelling Motif-12" section. The loop connecting Motif-3 and Motif-12 was modeled using the same loop modeling principles used in that section). Figure 4 (a), (b) show Motif-123 along with predicted residue-residue contacts between Motif-3 and neighboring SSES:- 7H, 1E.

Since, there is ambiguity here, we generated thousands of additional conformations of Motif-3 with respect to Motif-12, by sampling randomly in the phi-psi space of the loop residues connecting Motif-12 and Motif-3. Incidentally, Motif-123 remains of lowest energy (energy computed using the CHARMM force field[27] implementation in NAMD[28]) among this large set of additionally generated conformations (excluding structures close in geometry to Motif-123). However, we did find another NTD-hXPB structure, via this sampling process, which was of comparable energy to Motif-123. We call that as Motif-123-2 (shown in Figure 4 (c), (d)). Despite generating a large set of additional conformations, at this point, we do not claim that we have been able to exhaustively sample the conformation space of Motif-3 wrt Motif-12. Hence, we call models developed in this paper, Motif-123 and Motif-123-2, as “near-complete” instead of “complete” (We are very confident though that our models accurately capture the actual fold of NTD-hXPB).

Finally, after the above steps, we get two decoys/models for NTD-hXPB:- Motif-123 and Motif-123-2.

### Justifications Based on Simple Probability Analysis

Please refer to Supplementary Note 1, where we provide simple probability analysis (based on reasonable assumptions) to justify the key steps in the NTD-hXPB modelling protocol, that we just outlined.

### Modeling NTD-mXPB (N-terminal domain of XPB in MTB / TB-Pathogen)

Earlier works[8,29] have revealed homologue of NTD-hXPB in the TB pathogen - mycobacterium tuberculosis (MTB), which we call NTD-mXPB. The NTD-mXPB sequence is residues 1-120 of MTB XPB (http://www.uniprot.org/uniprot/O53873). In Supplementary Figure 4 we show results of sequence alignment between NTD-hXPB and NTD-mXPB (23% identity and 40% similarity). Supplementary Figure 5 shows secondary structure predictions for NTD-mXPB. As expected, they are similar to that of NTD-mXPB which is shown in Supplementary Figure 1. We also observed similar predicted contact residue profiles between them (details not presented in this paper). Finally, here we derive models for NTD-mXPB, from the derived NTD-mXPB models (Motif-123 & Motif-123-2), using the traditional homology modeling method. Figure 6 shows one of the two derived models, being compared with it's human homologue.

### Structure of NTD-hXPB bound to DNA

The binding geometry of a DNA binding helix-turn-helix (DB-HtH) motif with respect to a DNA strand is not arbitrary, but highly constrained. It can be determined with just one degree of freedom with a narrow upper and lower bound[30]. Hence we used the DB-HtH binding geometry parameters from PDB:6CRO to derive an approximate binding conformation of NTD-hXPB, via it’s DB-HtH, onto a generic DNA strand (also seen in Figure 5).

### Comparison with Predicted Models from Existing Automated Ab-Initio Tools

There are three well established automated ab-Intio structure prediction tools that we are aware of: ROSETTA/GREMLIN[31], QUARK[32], EVFOLD[33]. These use in-build contact residue prediction protocols. These three tools were also used to model NTD-hXPB (generated models uploaded at http://cs.stanford.edu/~mitul/research/ntd_xpb/, along with ones developed here). They all yielded distinct folds (shown in Supplementary Figure 6). We believe all these three methods fail to capture the actual structure of NTD-hXPBS. Some of the reasons are as follows. The model from EVFOLD is quite disordered as seen in Supplementary Figure 6 (c). ROSETTA/GREMLIN and QUARK report "parallel" pairing of sheets SSES(2):2E and SSES(4):4E, while we report "anti-parallel" pairing. We believe "anti-parallel" is the correct pairing type as, unlike ROSETTA and QUARK, we used a third party tool (BBCONTACTS) which is specifically designed and developed to detect beta sheet pairings with high accuracy. BBCONTACTS has been benchmarked on recent datasets and outperforms other sheet pairing methods by very significant margin[22]. QUARK also failed to capture the geometry of the helix-turn-helix DNA binding domain.

### TM-Scores of the Models

Popular and established structure modeling tool QUARK uses TM-score[34] to evaluate the quality of predicted 3D models for protein sequences. TM-score has a value in (0,1], where 1 indicates perfect match with the actual structure. We believe models developed in the paper can have TM-score anywhere in the range (0, 0.5). That is, at the very least, we expect our models to accurately capture the topology of the actual NTD-hXPB fold.

## RESULTS & DISCUSSIONS

## Role Of NTD-hXPB In Protein Synthesis and DNA Damage Repair

### XPB and TFIIH Anchor onto DNA via NTD-hXPB

We propose that the role of NTD-hXPB, in transcription (first step of protein synthesis) and DNA damage repair (specifically, NER), is to bind onto DNA which in turn provides anchoring to TFIIH and also to rest of XPB. We discuss this in details in following sections.

### Interactions between NTD-hXPB and p52

"Co-laborer" protein p52 (http://www.uniprot.org/uniprot/Q92759; subunit of human Transcription factor II H or TFIIH) interacts (forms contacts) with XPB (also part of TFIIH) in humans, via NTD-hXPB. These interactions/contacts are *essential* for successful progression of transcription and DNA damage repair (specifically, nucleotide excision repair or NER) in humans[3,35]. An outcome of p52 and NTD-hXPB interactions is anchoring of TFIIH onto XPB via the contacts/interactions. From [3], one might suspect the existence of p52 interaction site around residue F99. This possibility is further strengthened by our modeling outcomes in this paper, which shows F99 as an *exposed* hydrophobic residue and also a highly *conserved* residue (Figure 5). Recall, in the protein world, *conserved exposed* residues are ones that are very likely to participate in protein-protein interactions[26,36]. Moreover, this same potential p52 interaction site is poorly conserved in MTB (Figure 6: "region B"). This is expected as p52 interactions are suspected to be absent in MTB. This is because homologue of p52 has not been yet found in bacteria/MTB[8], and hence, in the first place there may not be a protein corresponding to p52 in bacteria. Overall, all these indicate that the region around F99, in our NTD-hXB model, (marked in Figure 5 and also in Figure 6 as "B") is indeed a potential p52 binding site in NTD-hXPB.

### Interactions between NTD-hXPB and DNA

Now, NTD-hXPB has a DNA binding domain (DBD) whose DNA binding region is exposed (as seen in Figure 5, which also shows a DNA bound conformation of NTD-hXPB). Hence, potentially, NTD-hXPB anchors onto DNA via it's DBD. We also know that p52 (and hence TFIIH minus NTD-hXPB; recall p52 is subunit of TFIIH) anchors onto NTD-hXPB, as discussed previously. As a result we can say that, potentially, p52 (and hence TFIIH minus NTD-hXPB), indirectly anchors onto DNA via NTD-hXPB.

### Relation of NTD-hXPB to rest of XPB

Also, via contact residue predictions (using CCMPRED/RAPTORX), we did not find any predicted interactions between NTD-hXPB and rest of XPB. In-fact the residue segment of XPB (201-314; Figure 8) that connects NTD-hXPB to the helicase region (residues 315-700; Figure 8) on XPB was found to be disordered via SSE prediction tools JPRED/PSIPRED/RAPTORX. (Figure 8 is reproduced from [8]). Hence NTD-hXPB potentially provides “anchoring” to rest of XPB. The latter being “tethered” to NTD-hXPB by the predicted disordered residues:201-314 between them. NTD-hXPB generates “anchoring” by binding to the DNA via it’s DNA binding motif. After being indirectly anchored to the DNA via NTD-hXPB, the rest of XPB can then perform it’s job of helicase activity on the DNA (*i.e.*, unwinding the DNA in both transcription and nucleotide excision repair). This is partly illustrated in Figure 7.

Summary of NTD-hXPB’s Role

So, to summarize the role of NTD-hXPB in the overall process of transcription and DNA damage repair (specifically NER): p52 anchors onto XPB via it’s contacts with NTD-hXPB. In turn, NTD-hXPB anchors onto the DNA via it’s DNA binding motif. This results in indirect anchoring of TFIIH onto the DNA via p52 and NTD-hXPB. NTD-hXPB also keeps the rest of XPB (especially it’s helicase) tethered via the disordered residue segment connecting them. This in turn keeps XPB indirectly anchored to DNA (via the binding of NTD-hXPB onto DNA) as the former performs helicase activity on the DNA (i.e., unwinding the DNA in transcription and nucleotide excision repair). The role of NTD-hXPB is partly illustrated in Figure 7.

The Second Binding Site of p52 on NTD-hXPB

p52 has a second “unknown” binding site on NTD-hXPB, as per [35]. We suspect it is in the region marked “C” in Figure 6 (a). This is because that region is poorly conserved in MTB but highly conserved in human (as seen in the figure), as was the case for the first potential p52 binding site (region “B”). We pointed out earlier that a potential p52 binding site is expected to be a highly conserved exposed region in human, but the corresponding region in MTB is likely to be poorly conserved. This is because p52, and hence, p52 interactions are believed to be absent in MTB.

## Directly Involvement in Diseases

XP-B

Heterozygous mutation F99S in NTD-hXPB is implicated in the xeroderma pigmentosum complementation group B (XP-B) - a rare, autosomal recessive disease[10]. The mutation leads to decreased: (a) DNA repair rates, (b) levels of XPB protein, and (c) basal transcription activity. The patients can experience extreme sensitivity to sunlight, pigmentation abnormalities, and skin cancer. Reduction of transcription activity in the patients can cause dwarfism, neuromyelination defects, deafness, and impaired sexual development (http://www.omim.org/entry/610651). Fewer than 40% of patients with the XP disease survive beyond the age of twenty.

Our modeling of NTD-hXPB reveals that the residue 99 is hydrophobic (Phe) and exposed (Figure 5/6). Phe, in general, has high propensity to form protein-protein contacts[36]. In-fact Phe when mutated

Ser is significantly less likely to participate in protein-protein interactions[36–38]. Hence residue F99 is a potentially a “functional” residue. From [3] also, one might suspect that F99 interacts with the “co-labourer” protein p52 in a way critical to both - the basal transcription activity as well as nucelotide excision repair (NER). Hence mutation of F99 can lead to serious functional loss in XPB potentially leading to diseases.

Statistical analysis of residue conservation (using CONSURF) reveals F99 as being highly conserved, apart from being exposed, in human (but not in MTB, as it seemingly lacks the homologue for the human XPB interaction partner p52; Figure 6). This further confirms F99's potential critical role in the functionality of the human transcription/DNA-repair factor XPB, which would go wrong if F99 is mutated.

XP-CS

XP-B is also known to co-occur with cockayne syndrome (CS) - another rare autosomal recessive. In XP-CS patients, one can additionally observe neurodegenerative disorders characterized by growth failure, impaired development of the nervous system, eye disorders and premature aging (https://en.wikipedia.org/wiki/Cockayne_syndrome).

TTD2

Homozygous mutation T119P in NTD-hXPB is implicated in another rare autosomal recessive disorder known as Trichothiodystrophy-2 (TTD2). http://www.omim.org/entry/616390 describe the condition as: “patients have brittle, sulfur-deficient hair … TTD patients display a wide variety of clinical features, including cutaneous, neurologic, and growth abnormalities. Common additional clinical features are ichthyosis, intellectual/developmental disabilities, decreased fertility, abnormal characteristics at birth, ocular abnormalities, short stature, and infections. There are both photosensitive and non-photosensitive forms of the disorder.”

In our derived model for NTD-hXPB, we notice T119 (marked in Figure 5/6 and Supplementary Figure 1) lies at a very short 1-3 residue gap between the fourth SSE (a sheet) and the fifth SSE (a helix). The very short residue gap between the two SSEs indicates that there is not much joint flexibility between the them. This is also the junction of Motif-1 and Motif-2. Since amino acid Proline has a constrained dihedral *phi,* T119P mutation is likely to alter the 3D orientation of Motif-2 with respect to Motif-1. This can result in the loss of structural integrity of NTD-hXPB and hence impairing the role of NTD-hXPB in the process of transcription and DNA repair. This could be the cause of disease/disorders arising from T119P mutation in NTD-hXPB.

Statistical analysis of residue conservation (using CONSURF) reveals T119 as a highly conserved (Figure 6) residue in human (as well as in MTB - unlike F99), further confirming it’s potential critical role in the functionality of the human transcription/DNA-repair factor XPB in both human and bacteria MTB. Hence mutation of T119 can cause “derailing” of XPB functionality leading to diseases.

## Human Homologue (NTD-hXPB) vs TB-Pathogen Homologue (NTD-mXPB)

Loss of p52 binding affinity

In the earlier sections "Interactions between NTD-hXPB and p52" and "The Second Binding Site of p52 on NTD-hXPB" we discussed how the "functionally critical" p52 interactions are potentially lost in the MTB homologue of NTD-hXPB. This is the major alteration of functionality between NTD-hXPB and it's homologue in the MTB/TB-pathogen, that we observe/confirm via residue conservation statistics.

Potential Therapeutic Target for Tuberculosis Disease (TB)

We also observe that a loop region, residues 34-36, is highly conserved and exposed in NTD-mXPB but not significantly conserved in NTD-hXPB (as seen in region "A" in Figure 6). We discussed earlier that such exposed and conserved regions are potential functional/interaction sites[26,36]. Hence this region in NTD-mMTB could be potential therapeutic target for TB.

So far, our discussions were centered around the first model of NTD-hXPB – motif-123. The conclusions also hold for the second model – motif-123-2.

## CONCLUSIONS AND FUTURE DIRECTIONS

In this paper we derive near-complete 3D models for NTD-hXPB - the N-terminal domain of the human XPB *aka* xeroderma pigmentosum type B protein. The results presented in this paper are very significant because:-

- NTD-hXPB plays a critical role in transcription - the first step in **synthesis of proteins** (proteins perform almost all tasks in the cell and cell is the building block of life) and in **DNA damage repair** (specifically nucleotide excision repair - NER). NER is in turn linked to cancer and aging.
- NTD-hXPB is part of XPB which, along with basal transcription machinery, are **prospective targets of anti-cancer therapies**.
- NTD-hXPB has a homologue in *mycobacterium tuberculosis* (MTB). The structure of the homologue, that we derive using the model for NTD-hXPB, could be potential therapeutic target for the infectious *tuberculosis (*TB) disease caused by MTB. According to WHO, **TB surpassed HIV as #1 infectious disease killer in 2014.**
- NTD-hXPB is implicated in the xeroderma pigmentosum complementation group B (XP-B) - a **rare, autosomal recessive disease**. The disease leads to reduced DNA repair rates and basal transcription activity. XP-B is known to co-occur with cockayne syndrome (CS) - another rare autosomal recessive. XP-CS patients can experience **neurodegenerative disorders** characterized by growth failure, impaired development of the nervous system, eye disorders and **premature aging**. NTD-hXPB is implicated in yet another rare autosomal recessive disorder known as Trichothiodystrophy-2 (TTD2). TTD patients can display a wide variety of clinical features, including cutaneous, neurologic, growth abnormalities, ichthyosis, intellectual/developmental disabilities, **decreased fertility**, abnormal characteristics at birth, ocular abnormalities, short stature, and infections.
- We used a novel ab-initio (due to lack of structural homology) semi-automated computational approach to model NTD-hXPB, based on recent successes in protein structure prediction. We used the computational route, because years of experimental research on XPB had only succeeded in determining the structure of XPB minus NTD-hXPB.
- Our computational approach, based on human judgements, is in line with prevailing theory that **computational abilities of humans** can be a very powerful ally of machines.
- Using developed models, we are able to propose **significant insights** into (a) the role of NTD-hXPB in the process of transcription and DNA damage repair (specifically NER) (b) the critical interactions of NTD-hXPB with another protein p52, and (c) **rare diseases associated with NTD-hXPB**.
- We also derive models of the structural homologue (NTD-mXPB) in the tuberculosis (TB) pathogen MTB. We compare NTD-hXPB and NTD-mXPB and propose explanations for alterations in functions between them.

- We also point to potential **<u>TB therapeutic target</u>** in the NTD-mXPB model.

**In the future,** we would like to:

- have the community use our novel computational protein structure prediction protocol, involving human, to determine structures of other protein sequences.
- perform “Crowd-Tasking” - Along the lines of FoldIt[12] and works of Luis von Ahn[13], we would like to have each one of available human across the planet (potentially billions) pick up a sequence with unknown structure (millions in http://us.expasy.org/sprot/) and use our structure prediction approach to determine it’s structure. This idea has the potential to significantly narrow the vast gap between available protein sequences (millions) and known protein structures (~100,000).
- investigate how the developed models for NTD-hXPB can help with the development of anti-cancer therapies[6,7].
- investigate how the developed models for NTD-hXPB can help with understanding and developing cures for associated diseases (XP-B, XP-CS, TTD2)
- investigate how the developed models for NTD-mXPB can help with understanding and developing cures for TB.
- seek further understanding into the mechanisms of transcription / DNA-repair (and hence it’s link to cancer & aging) using similar approaches. That is, first derive models of other components (i.e., proteins) involved in the mechanisms, then determine how the different components interact with each other, and so on. Our next target is the p52 protein - a “co-labourer” of XPB in transcription and DNA repair.

## ACCESSING THE MODELS

All the developed models have been uploaded to:- http://cs.stanford.edu/~mitul/research/ntd_xpb/

## ACKNOWLEDGMENT

The author is thankful to Dr. Marc Morais Lab (Sealy Center for Structural Biology at University of Texas Medical Branch at Galveston) for partly funding this work. This work has benefited from electronic discussions with: Dr. Jessica Andreani Feuillet (Johannes Soeding Lab), Stefan Seemayer (Johannes Soeding Lab), Sergey Ovchinnikov (David Baker Lab), Dr. Jinbo Xu (RAPTORX developer), Dr. Alexey Drozdetskiy (Barton Lab), and Dr. Bent Petersen (NETTURNP developer). Some of the computations in this work was performed at "Texas Advanced Computing Center – TACC" computing cluster (https://www.tacc.utexas.edu/).

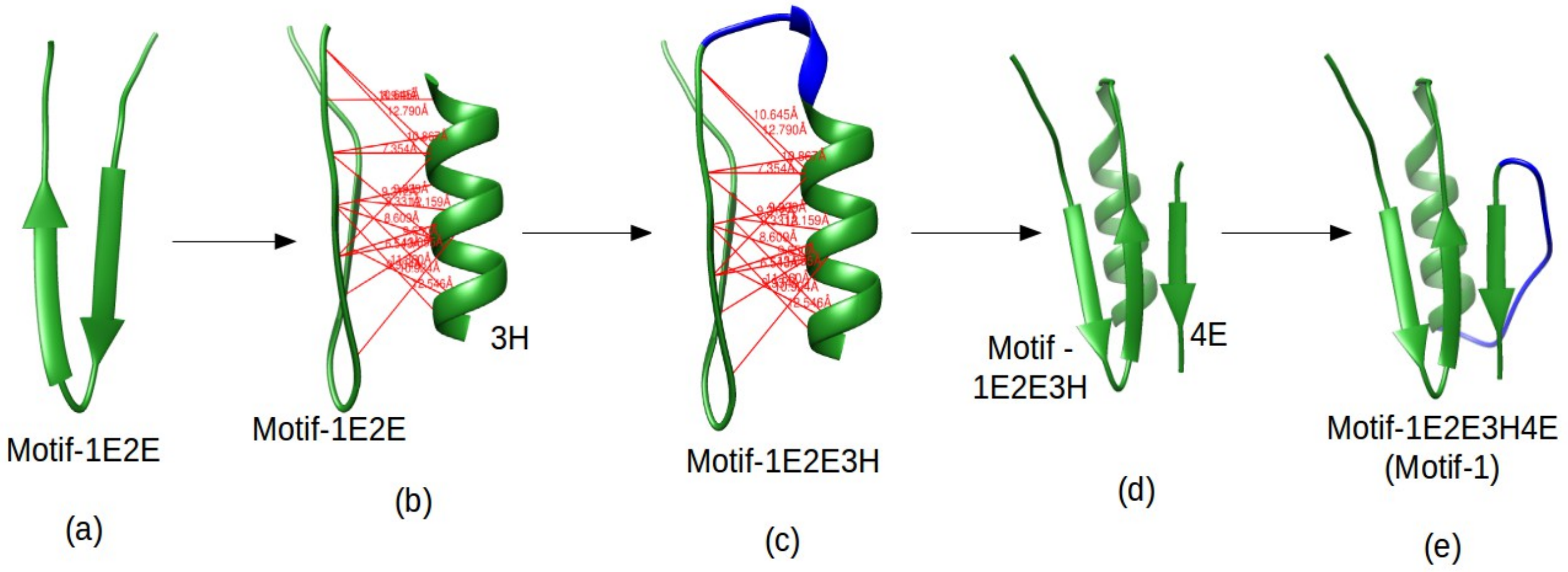

**Figure 1**: "Modeling Motif-1": (a) The first two SSEs (1E, 2E), forms a beta hairpin (Motif-1E2E) which was modelled using known properties of beta hairpin and sheet pairing predictions from BBCONTACTS. (b) Then the second SSE (3H), a helix, was placed by a human besides Motif-1E2E (shown on left from a different view than in (a)), such that most of the predicted residue-residue contacts were satisfied. (c) The loop connecting the third SSE and the previous SSE (region show in blue) was modelled using Modeller. (d) Then the fourth SSE (4E; shown on right) was placed, in the 3D space, as the third beta sheet to pair with the beta sheet pair Motif-1E2E (the one shown in (a)) using sheet pairing predictions from BBCONTACTS. This gives the Motif-1E2E3H4E or final Motif-1.

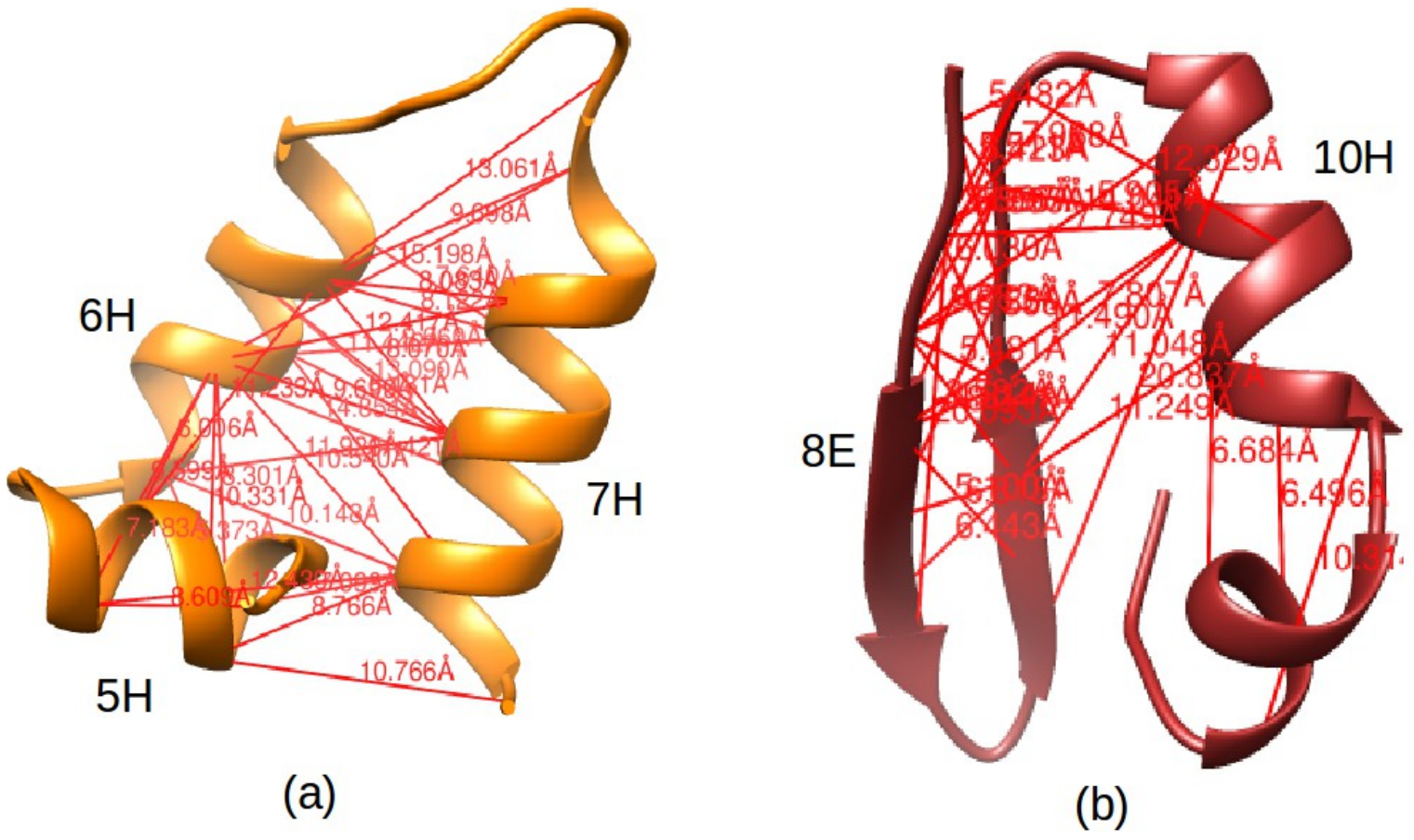

**<u>Figure 2</u>** (a): The final constructed Motif-2 with labeled (via line segments) predicted contact residue pairs. The small helix, closer to figure bottom edge, is the first helix of the helix-turn-helix DNA binding motif. (b): Modelled Motif-3 with labeled (via line segments) predicted contact residue pairs.

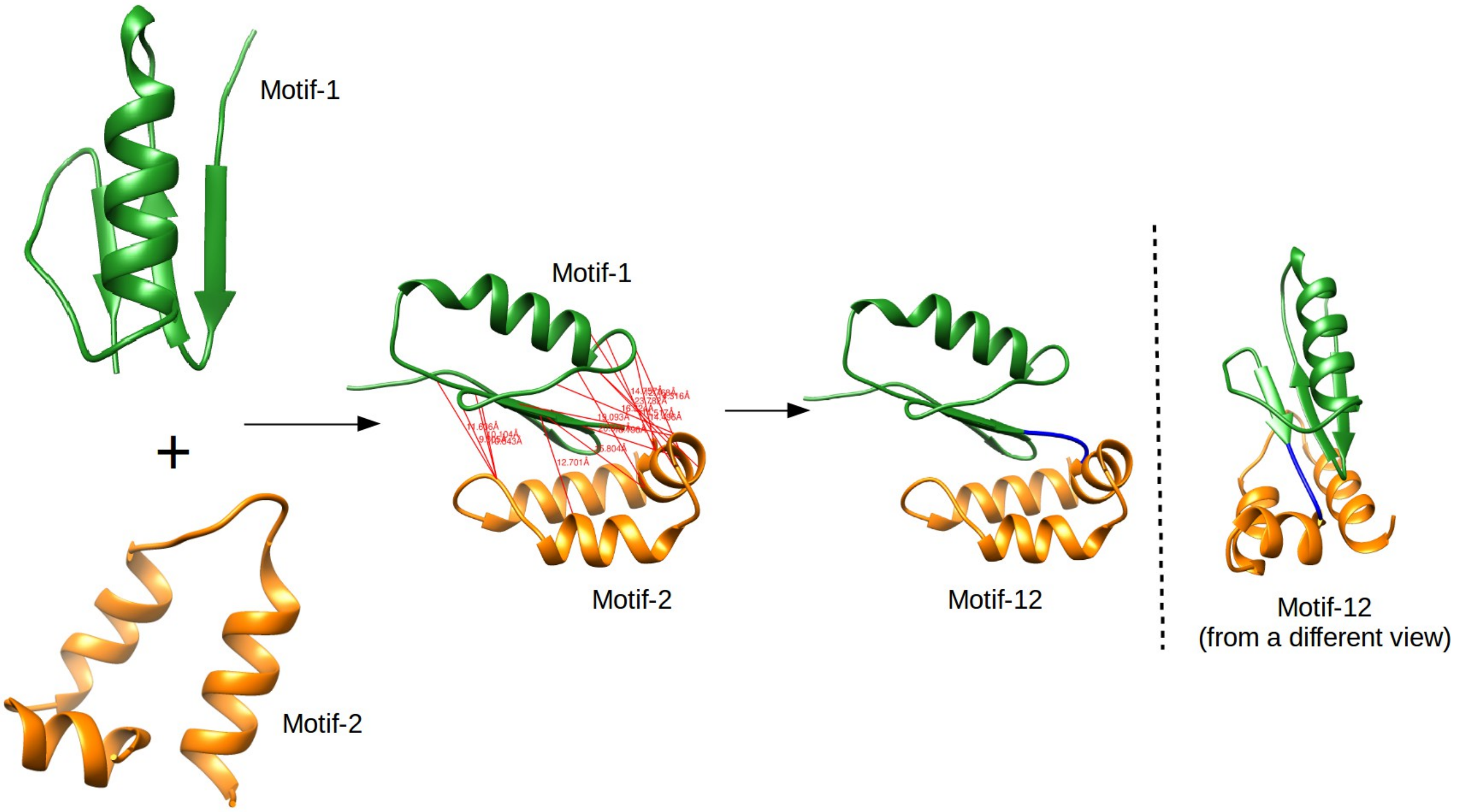

**Figure 3**: Steps of Motif-12 building illustrated. Shown (red lines) are predicted residue-residue contacts between Motif-2 and the neighboring beta-sheet region of Motif-1.

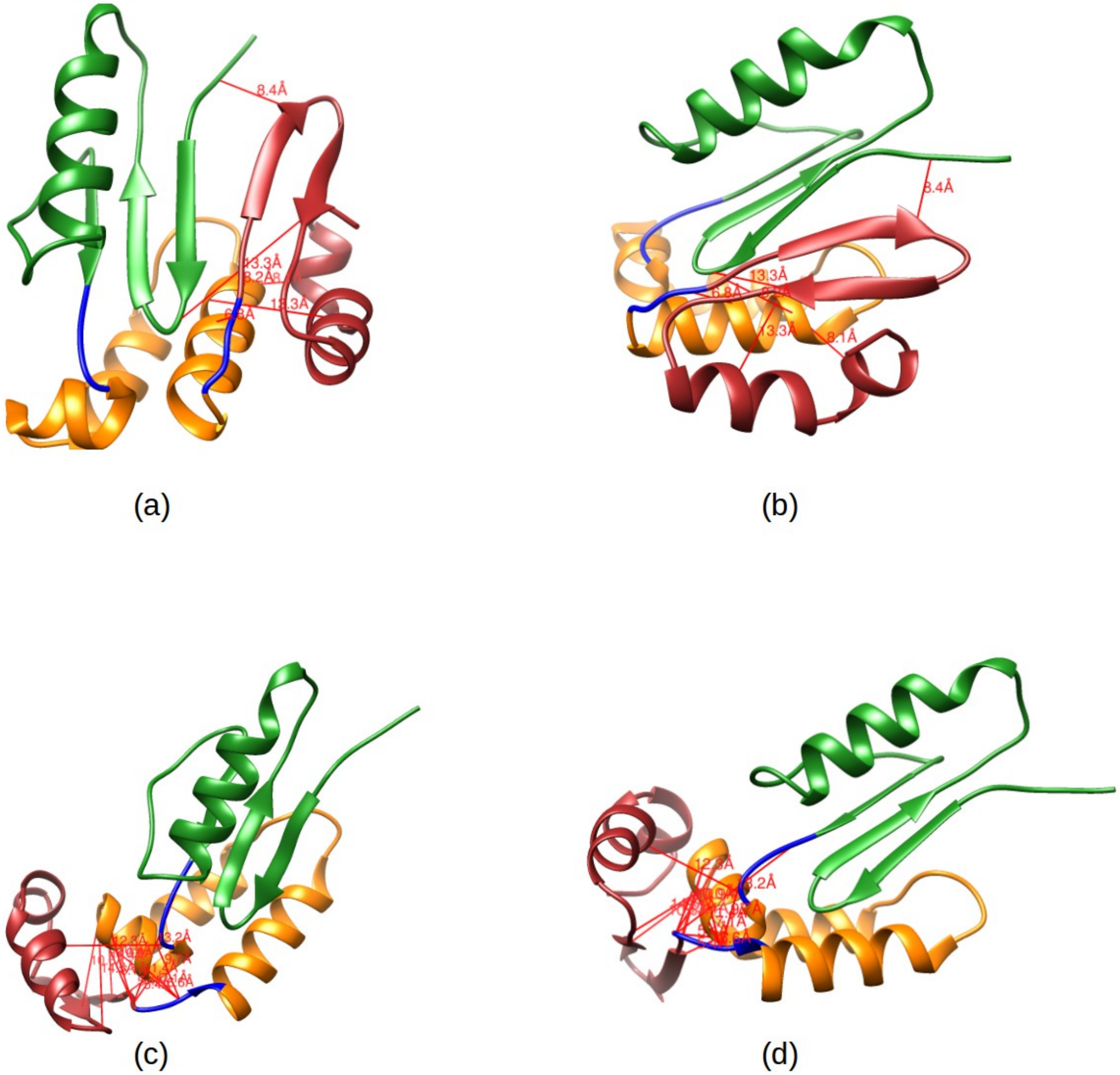

**<u>Figure 4</u>** (a): Motif-123 (a final NTD-hXPB model) along with predicted (shown via line labels) residue-residue contacts between Motif-3 (brown portion) and neighboring SSES (7H, 1E) (b): Motif-123 in (a) shown from a different view.

(c): Motif-123-2 - second model for NTD-hXPB, obtained via sampling of Motif-3 around Motif-12. Also shown are predicted residue-residue contacts between Motif-3 and the nearby SSE - the first helix of the HtH DNA binding motif. (d): Motif-123-2 in (c) shown from a different view

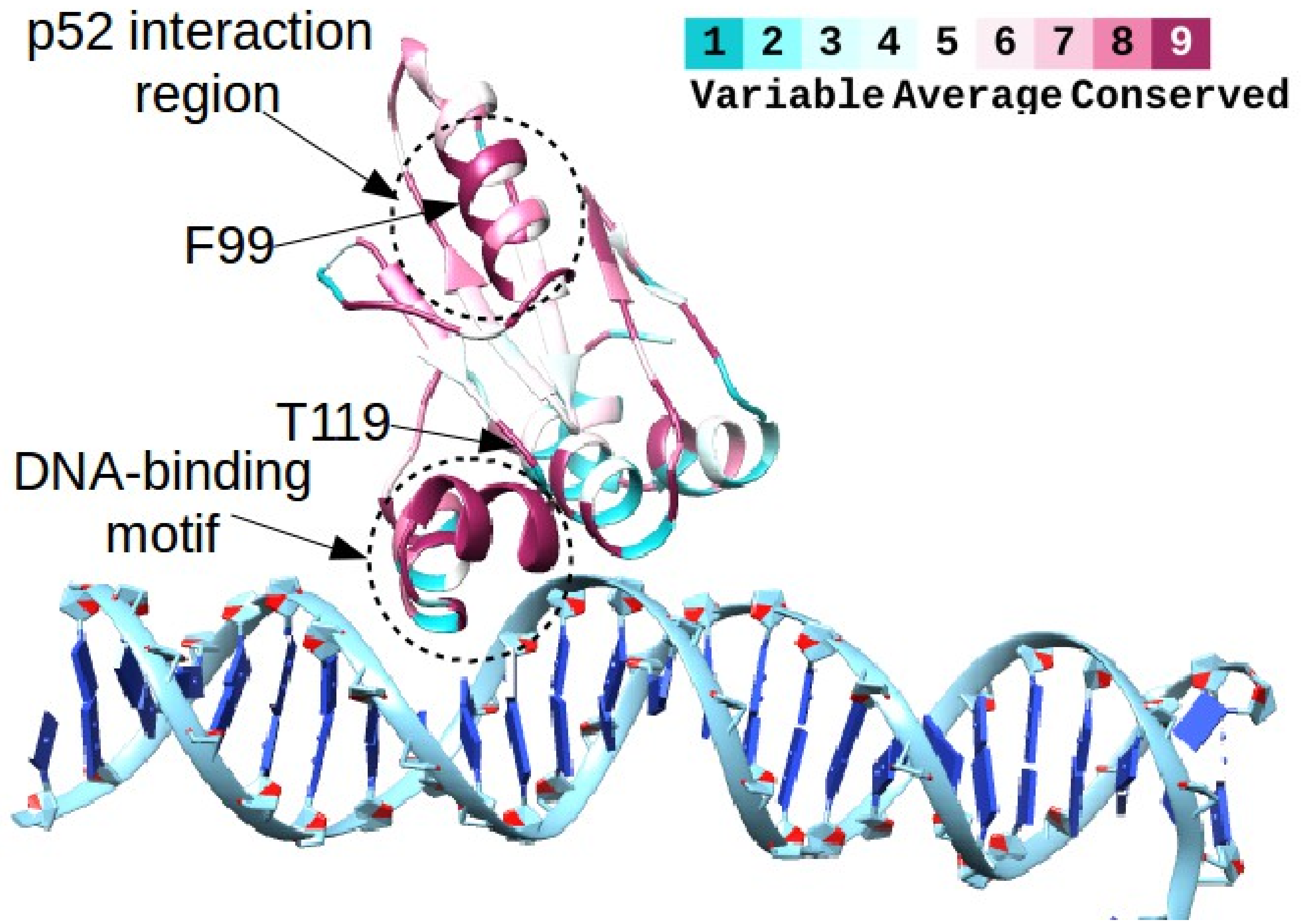

**Figure 5**: A DNA bound conformation of NTD-hXPB. NTD-hXPB is coloured as per conservation statistics on it's residues computed by CONSURF. Top right portion shows the colouring code. The potential p52 interaction site/region, around residue F99, is marked by dotted circle. The site is a "highly conserved" region as per the colouring code.

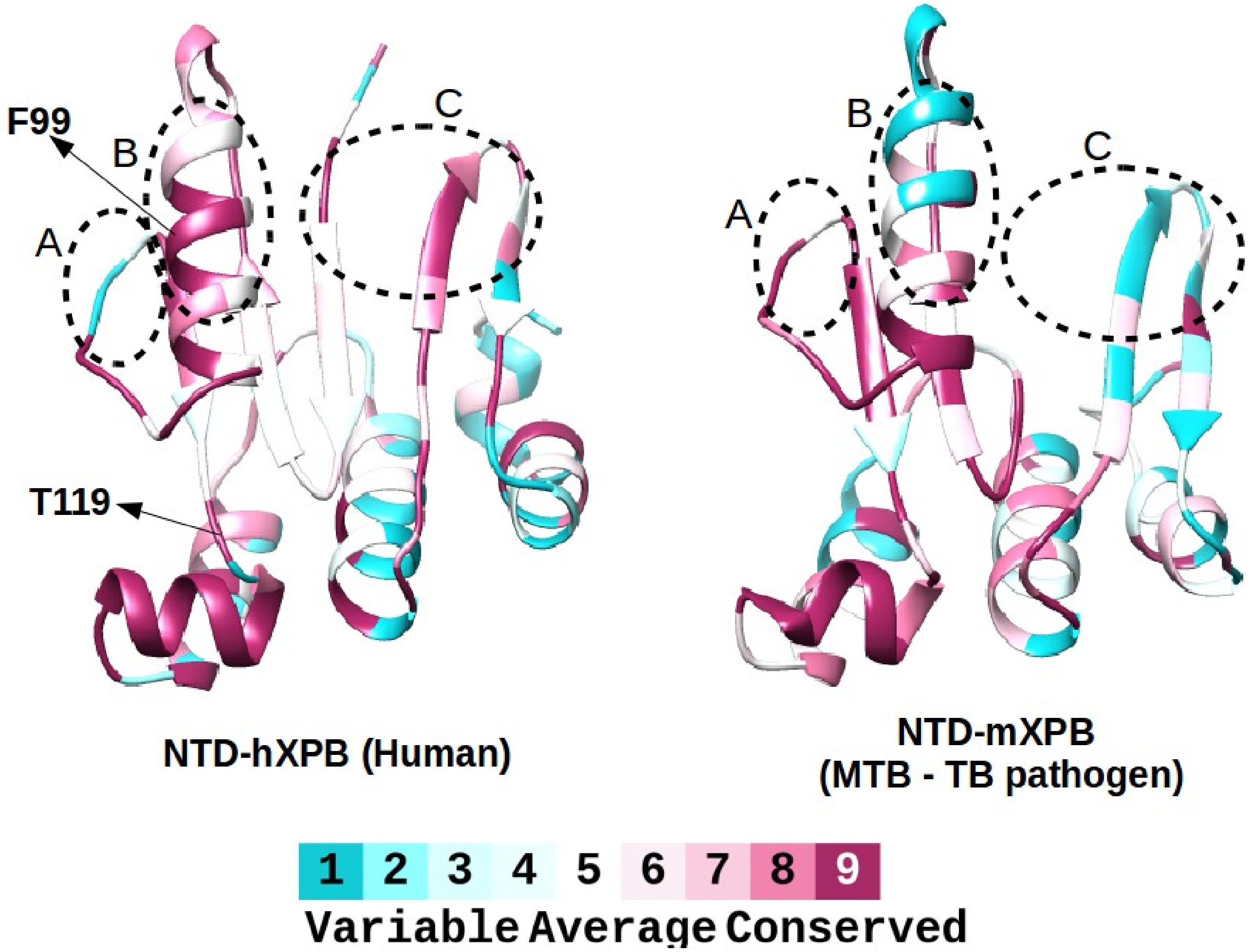

**Figure 6**: NTD-hXPB (structure on left; from human) vs NTD-mXPB (structure on right; from MTB – TB pathogen). They are coloured as per conservation statistics on their residues computed by CONSURF. Bottom portion shows the colouring code. Three regions ("A", "B", "C") are marked in NTD-hXPB and it's homologue NTD-mXPB. "B" corresponds to suspected p52 binding site in NTD-hXPB. "C" corresponds to potential second p52 binding site in NTD-hXPB. "A" in NTD-mXPB corresponds to potential target for TB-therapeutics.

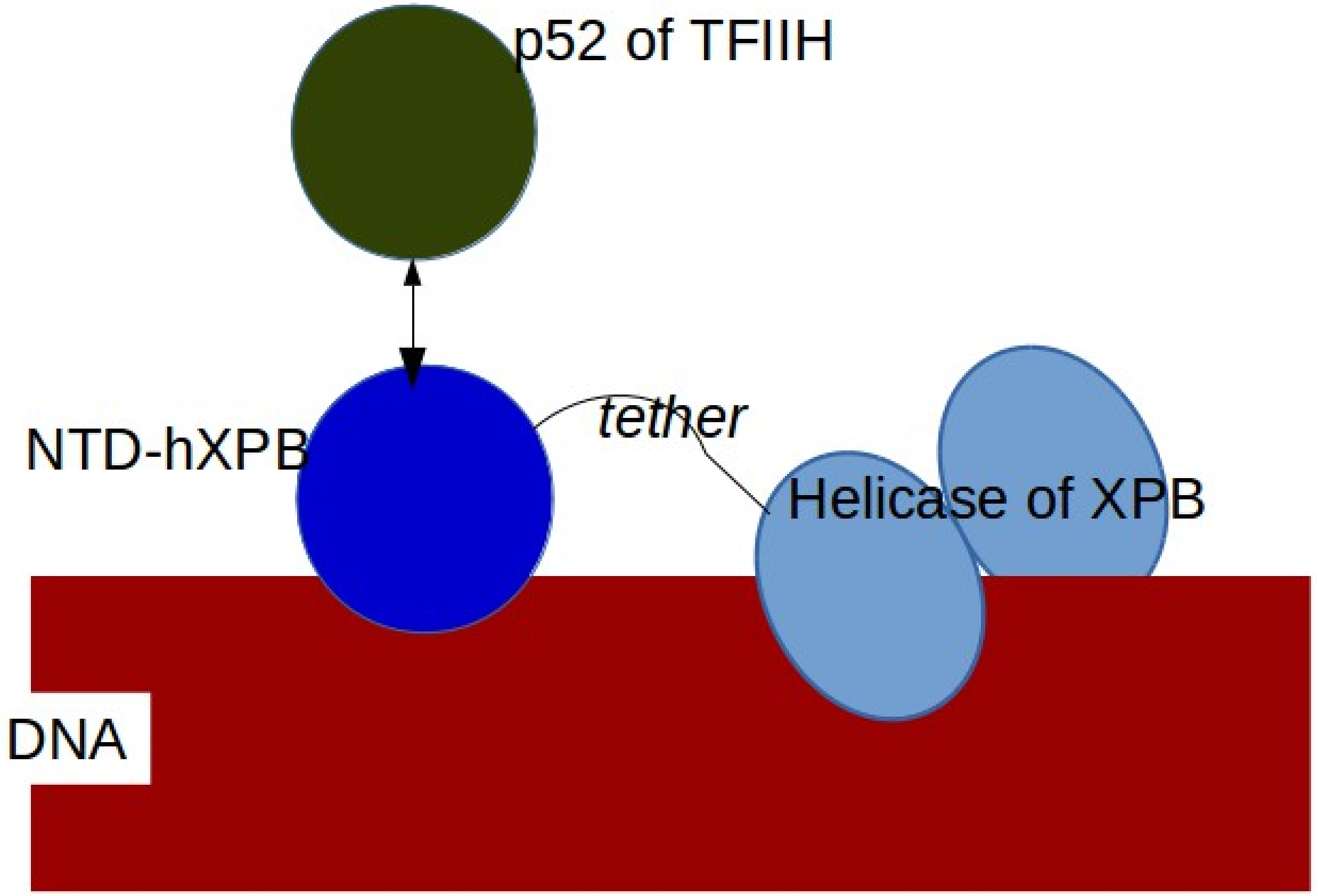

**Figure 7**: The potential role of NTD-hXPB shown in a cartoon representation. NTD-hXPB, the N-terminal domain of human XPB, “anchors” onto the DNA via its DNA binding domain. The disordered chain of amino acid between it and the helicase domain of XPB keeps the helicase “tethered” to NTD-hXPB. The p52 of THIIH indirectly “anchors” onto DNA via the already anchored NTD-hXPB.

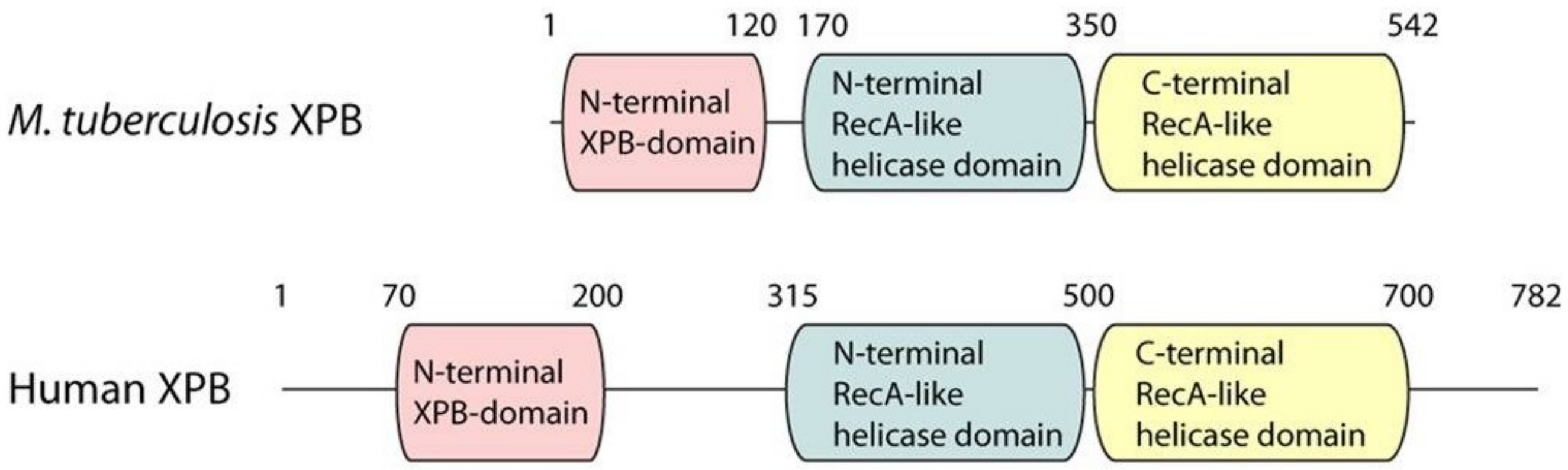

**Figure 8**: Figure reproduced from an earlier work (see main text), **showing domain boundaries for protein XPB in MTB (top) and human (bottom).**

## SUPPLEMENTARY INFORMATION

### Table of Contents

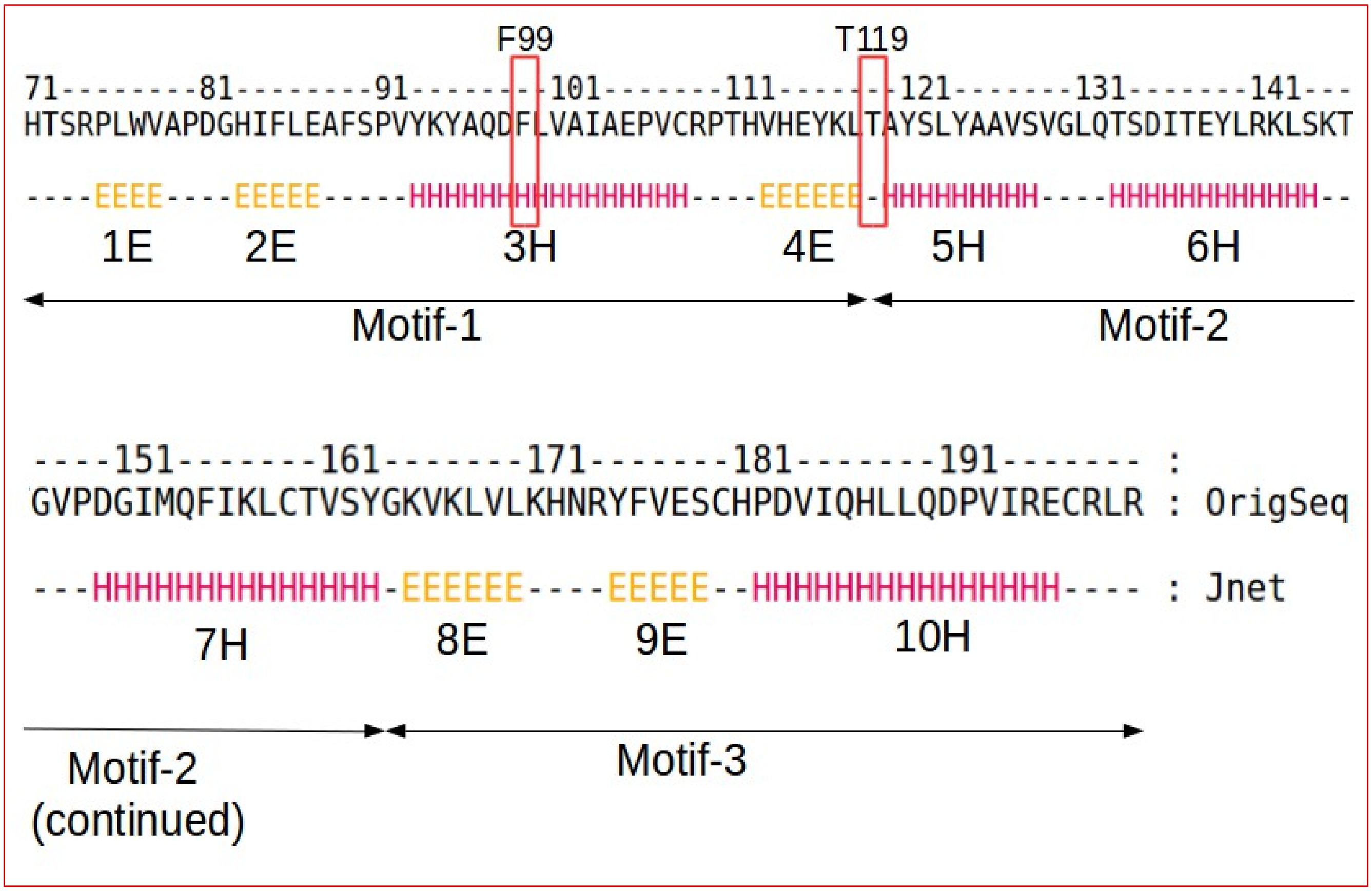

**<u>Supplementary Figure 1</u>**: Secondary structure element (SSE) predictions for the NTD-hXPB sequence (residues 71-200 of human XPB: http://www.uniprot.org/uniprot/P19447). We have confirmed that these SSE predictions from JPRED are consistent with SSE predictions from other two known SSE predictors: PSIPRED and RAPTORX. There are ten predicted SSEs (sheet:1E, sheet:2E, helix:3H, sheet:4E, helix:5H, helix:6H, helix:7H, helix:8E, helix:9E, helix:10H). For easy future reference, we store these ten predicted SSEs as a sequence SSES:- (1E, 2E, 3H, 4E, 5H, 6H, 7H, 8E, 9E, 10H). **Residues which cause rare disease, upon mutation, are shown: 99F, 119T**.

| #identifier | diversity | direction | viterbiscore | indexpred | state | res1 | res2 | |
|---|---|---|---|---|---|---|---|---|
| query | 0.17 | Antiparallel | 0.707922 | 1 | first | 42 | 18 | |
| query | 0.17 | Antiparallel | 0.707922 | 1 | internal | 43 | 17 | |
| query | 0.17 | Antiparallel | 0.707922 | 1 | internal | 44 | 16 | Predicted pairing of 2E, 4E |
| query | 0.17 | Antiparallel | 0.707922 | 1 | internal | 45 | 15 | |
| query | 0.17 | Antiparallel | 0.707922 | 1 | internal | 46 | 14 | |
| query | 0.17 | Antiparallel | 0.707922 | 1 | internal | 47 | 13 | |
| query | 0.17 | Antiparallel | 0.707922 | 1 | last | 48 | 12 | |
| query | 0.17 | Antiparallel | 0.705088 | 2 | first | 105 | 100 | |
| query | 0.17 | Antiparallel | 0.705088 | 2 | internal | 106 | 99 | Predicted pairing of 8E, 9E |
| query | 0.17 | Antiparallel | 0.705088 | 2 | internal | 107 | 98 | |
| query | 0.17 | Antiparallel | 0.705088 | 2 | internal | 108 | 97 | |
| query | 0.17 | Antiparallel | 0.705088 | 2 | internal | 109 | 96 | |
| query | 0.17 | Antiparallel | 0.705088 | 2 | last | 110 | 95 | |
| query | 0.17 | Antiparallel | 0.562766 | 3 | first | 13 | 9 | |
| query | 0.17 | Antiparallel | 0.562766 | 3 | internal | 14 | 8 | Predicted pairing of 1E, 2E |
| query | 0.17 | Antiparallel | 0.562766 | 3 | internal | 15 | 7 | |
| query | 0.17 | Antiparallel | 0.562766 | 3 | internal | 16 | 6 | |
| query | 0.17 | Antiparallel | 0.562766 | 3 | last | 17 | 5 | |
| query | 0.17 | Antiparallel | -4.425870 | 6 | first | 43 | 37 | |
| query | 0.17 | Antiparallel | -4.425870 | 6 | last | 44 | 36 | |
| query | 0.17 | Parallel | -4.938243 | 7 | first | 46 | 6 | |
| query | 0.17 | Parallel | -4.938243 | 7 | internal | 47 | 7 | |
| query | 0.17 | Parallel | -4.938243 | 7 | last | 48 | 8 | |
| query | 0.17 | Antiparallel | -6.517789 | 18 | first | 108 | 105 | |
| query | 0.17 | Antiparallel | -6.517789 | 18 | last | 109 | 104 | |

**Supplementary Figure 2**. Beta sheet pairing predictions from BBCONTACTS. Top three predictions, with, positive scores, are marked by red boxes. Actual residues 71-200 are renumbered to 1-130 here.

```
# For publication of results, please cite:
# NetTurnP - Neural Network Prediction of Beta-turns by Use of Evolutionary Information and Predicted Protein Sequence Features.
# Petersen B, Lundegaard C, Petersen TN (2010)
# PLoS ONE 5(11):e15079 doi:10.1371/journal.pone.0015079
#
# Column 1:     Amino acid
# Column 2:     Sequence name
# Column 3:     Amino acid number
# Column 4:     Prediction for Beta-turn
# Column 5:     Class assignment - "T" for Beta-turn
# Column 6:     Prediction for Beta-turn type A
# Column 7:     Class assignment - "A" for Beta-turn type I
# Column 8:     Prediction for Beta-turn type B
# Column 9:     Class assignment - "B" for Beta-turn type I'
# Column 10:    Prediction for Beta-turn type C
# Column 11:    Class assignment - "C" for Beta-turn type II
# Column 12:    Prediction for Beta-turn type D
# Column 13:    Class assignment - "D" for Beta-turn type II'
# Column 14:    Prediction for Beta-turn type E
# Column 15:    Class assignment - "E" for Beta-turn type IV
# Column 16:    Prediction for Beta-turn type F
# Column 17:    Class assignment - "F" for Beta-turn type VIII
# Column 18:    Prediction for Beta-turn type G
# Column 19:    Class assignment - "G" for Beta-turn type VIb
# Column 20:    Prediction for Beta-turn type H
# Column 21:    Class assignment - "H" for Beta-turn type VIa1
# Column 22:    Prediction for Beta-turn type I
# Column 23:    Class assignment - "I" for Beta-turn type VIa2
#
H  Sequence   71   0.792 T  0.792 A 0.363 . 0.647 C 0.571 D 0.637 E 0.605 F 0.278 . 0.348 . 0.406 .
T  Sequence   72   0.726 T  0.782 A 0.380 . 0.483 . 0.547 D 0.661 E 0.601 F 0.266 . 0.267 . 0.398 .
S  Sequence   73   0.711 T  0.794 A 0.354 . 0.430 . 0.572 D 0.631 E 0.541 F 0.600 G 0.536 H 0.475 .
R  Sequence   74   0.722 T  0.790 A 0.319 . 0.405 . 0.516 D 0.604 E 0.563 F 0.557 G 0.458 . 0.376 .
P  Sequence   75   0.413 .  0.308 . 0.171 . 0.233 . 0.245 . 0.497 . 0.579 F 0.541 G 0.422 . 0.320 .
L  Sequence   76   0.289 .  0.197 . 0.104 . 0.134 . 0.140 . 0.363 . 0.541 F 0.566 G 0.412 . 0.320 .
W  Sequence   77   0.247 .  0.152 . 0.089 . 0.114 . 0.097 . 0.375 . 0.555 F 0.151 . 0.168 . 0.273 .
V  Sequence   78   0.208 .  0.138 . 0.108 . 0.128 . 0.179 . 0.327 . 0.461 . 0.255 . 0.236 . 0.443 .
A  Sequence   79   0.784 T  0.782 A 0.223 . 0.251 . 0.492 . 0.518 E 0.437 . 0.298 . 0.431 . 0.576 I
P  Sequence   80   0.817 T  0.795 A 0.498 . 0.340 . 0.657 D 0.575 E 0.460 . 0.395 . 0.426 . 0.573 I
D  Sequence   81   0.818 T  0.805 A 0.480 . 0.359 . 0.680 D 0.570 E 0.492 . 0.402 . 0.434 . 0.546 I
G  Sequence   82   0.835 T  0.819 A 0.458 . 0.347 . 0.671 D 0.611 E 0.476 . 0.273 . 0.313 . 0.339 .
H  Sequence   83   0.536 T  0.318 . 0.404 . 0.241 . 0.551 D 0.494 . 0.418 . 0.211 . 0.160 . 0.231 .
I  Sequence   84   0.217 .  0.122 . 0.136 . 0.119 . 0.254 . 0.298 . 0.398 . 0.112 . 0.131 . 0.207 .
F  Sequence   85   0.157 .  0.094 . 0.098 . 0.103 . 0.092 . 0.237 . 0.336 . 0.119 . 0.125 . 0.219 .
L  Sequence   86   0.228 .  0.119 . 0.099 . 0.116 . 0.085 . 0.305 . 0.366 . 0.143 . 0.131 . 0.263 .
E  Sequence   87   0.365 .  0.289 . 0.130 . 0.215 . 0.160 . 0.408 . 0.404 . 0.300 . 0.207 . 0.442 .
A  Sequence   88   0.538 T  0.311 . 0.208 . 0.327 . 0.254 . 0.539 E 0.571 F 0.509 G 0.300 . 0.459 .
F  Sequence   89   0.534 T  0.323 . 0.249 . 0.329 . 0.329 . 0.571 E 0.567 F 0.692 G 0.656 H 0.631 I
S  Sequence   90   0.705 T  0.708 A 0.248 . 0.445 . 0.365 . 0.604 E 0.582 F 0.689 G 0.608 H 0.630 I
P  Sequence   91   0.686 T  0.725 A 0.228 . 0.383 . 0.329 . 0.585 E 0.549 F 0.598 G 0.499 H 0.459 .
V  Sequence   92   0.624 T  0.703 A 0.129 . 0.276 . 0.220 . 0.483 . 0.361 . 0.498 . 0.634 H 0.470 .
Y  Sequence   93   0.604 T  0.713 A 0.093 . 0.362 . 0.153 . 0.435 . 0.348 . 0.133 . 0.330 . 0.279 .
K  Sequence   94   0.310 .  0.335 . 0.093 . 0.203 . 0.093 . 0.329 . 0.279 . 0.115 . 0.263 . 0.226 .
Y  Sequence   95   0.248 .  0.299 . 0.088 . 0.175 . 0.078 . 0.291 . 0.243 . 0.090 . 0.285 . 0.225 .
A  Sequence   96   0.267 .  0.286 . 0.096 . 0.223 . 0.087 . 0.283 . 0.196 . 0.084 . 0.116 . 0.196 .
Q  Sequence   97   0.190 .  0.198 . 0.102 . 0.119 . 0.104 . 0.242 . 0.109 . 0.082 . 0.110 . 0.193 .
D  Sequence   98   0.179 .  0.193 . 0.091 . 0.104 . 0.099 . 0.256 . 0.098 . 0.077 . 0.124 . 0.190 .
F  Sequence   99   0.176 .  0.205 . 0.087 . 0.099 . 0.091 . 0.257 . 0.103 . 0.085 . 0.124 . 0.186 .
L  Sequence   100  0.201 .  0.251 . 0.086 . 0.091 . 0.086 . 0.266 . 0.118 . 0.097 . 0.127 . 0.189 .
V  Sequence   101  0.217 .  0.275 . 0.083 . 0.099 . 0.080 . 0.290 . 0.135 . 0.095 . 0.125 . 0.194 .
A  Sequence   102  0.260 .  0.325 . 0.082 . 0.096 . 0.075 . 0.308 . 0.124 . 0.090 . 0.128 . 0.199 .
I  Sequence   103  0.294 .  0.345 . 0.087 . 0.106 . 0.087 . 0.334 . 0.249 . 0.103 . 0.150 . 0.204 .
A  Sequence   104  0.321 .  0.318 . 0.099 . 0.114 . 0.092 . 0.374 . 0.359 . 0.335 . 0.243 . 0.231 .
E  Sequence   105  0.351 .  0.356 . 0.100 . 0.128 . 0.093 . 0.417 . 0.419 . 0.293 . 0.243 . 0.238 .
P  Sequence   106  0.421 .  0.393 . 0.109 . 0.173 . 0.098 . 0.478 . 0.502 F 0.336 . 0.247 . 0.253 .
V  Sequence   107  0.414 .  0.355 . 0.138 . 0.204 . 0.114 . 0.507 E 0.474 . 0.339 . 0.249 . 0.263 .
C  Sequence   108  0.471 .  0.380 . 0.251 . 0.269 . 0.223 . 0.519 E 0.465 . 0.393 . 0.310 . 0.304 .
R  Sequence   109  0.514 T  0.494 . 0.270 . 0.389 . 0.467 . 0.526 E 0.426 . 0.418 . 0.349 . 0.317 .
```

**<u>Supplementary Figure 3</u>**. Shown above is partial predictions of beta turn types from NETTURNP. Marked by red box is the predictions for the first beta turn type for residues 79-82.

```
# Aligned_sequences: 2
# 1: EMBOSS_001
# 2: EMBOSS_001
# Matrix: EBLOSUM62
# Gap_penalty: 10.0
# Extend_penalty: 0.5
#
# Length: 132
# Identity:      30/132 (22.7%)
# Similarity:    53/132 (40.2%)
# Gaps:          14/132 (10.6%)
# Score: 86.5
#
#
#=======================================

EMBOSS_001         1 HTSRPLWVAPDGHIFLEAFSPVYKYAQDFLVAIAEPVCRPTHVHEYKLTA     50
                            :..|..:.||....:...|:..:...||....|.|||.|::|.
EMBOSS_001         1 -------MQSDKTVLLEVDHELAGAARAAIAPFAELERAPEHVHTYRITP     43

EMBOSS_001        51 YSLYAAVSVGLQTSDITEYLRKLSKTGVPDGIMQFIKLCTVSYGKVKLVL    100
                     .:|:.|.:.|.....:.:.|...|:..||..::..|......||:::||.
EMBOSS_001        44 LALWNARAAGHDAEQVVDALVSYSRYAVPQPLLVDIVDTMARYGRLQLVK     93

EMBOSS_001       101 K--HNRYFVESCHPDVIQHLLQDPVIRECRLR    130
                     .  |....| |....|::.:|::..|..
EMBOSS_001        94 NPAHGLTLV-SLDRAVLEEVLRNKKIAP----    120
```

**<u>Supplementary Figure 4</u>**. Alignment of the NTD-hXPB and NTD-mXPB sequences. NEEDLE (http://www.ebi.ac.uk/Tools/psa/) was used. Residues 71-200 of NTD-hXPB was renumbered to 1-130.

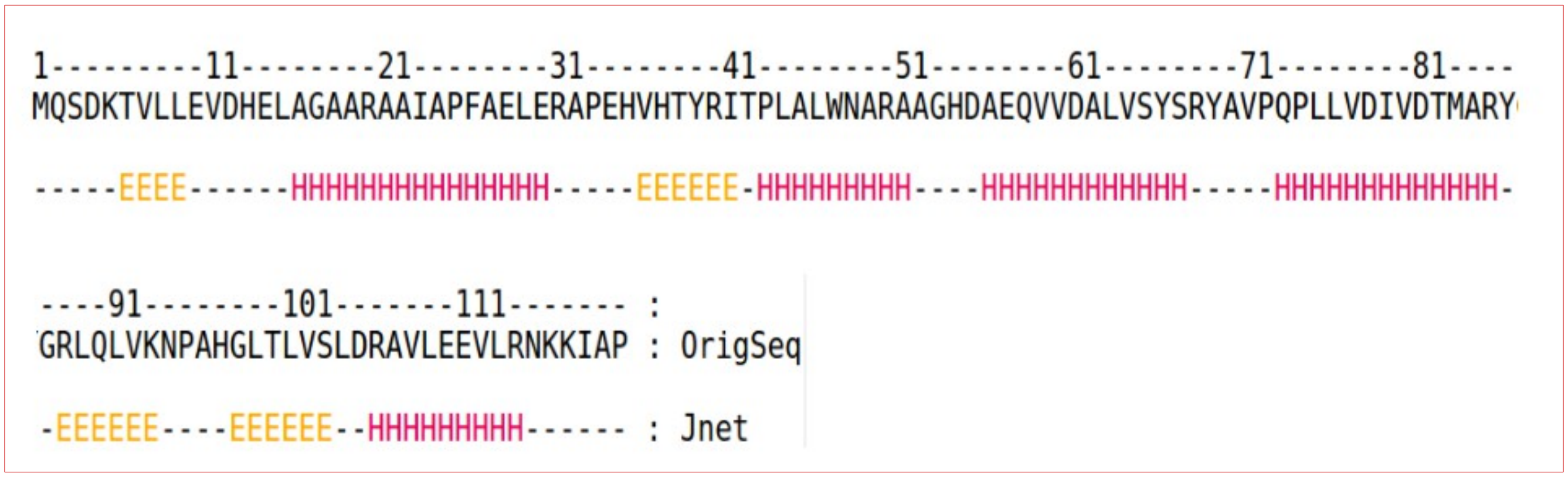

**<u>Supplementary Figure 5</u>**. Secondary structure element (SSE) predictions for the NTD-mXPB Sequence (residues 1-120 of MTB XPB: http://www.uniprot.org/uniprot/O53873). We have confirmed that these SSE predictions from JPRED are consistent with SSE predictions from other two known SSE predictors: PSIPRED and RAPTORX.

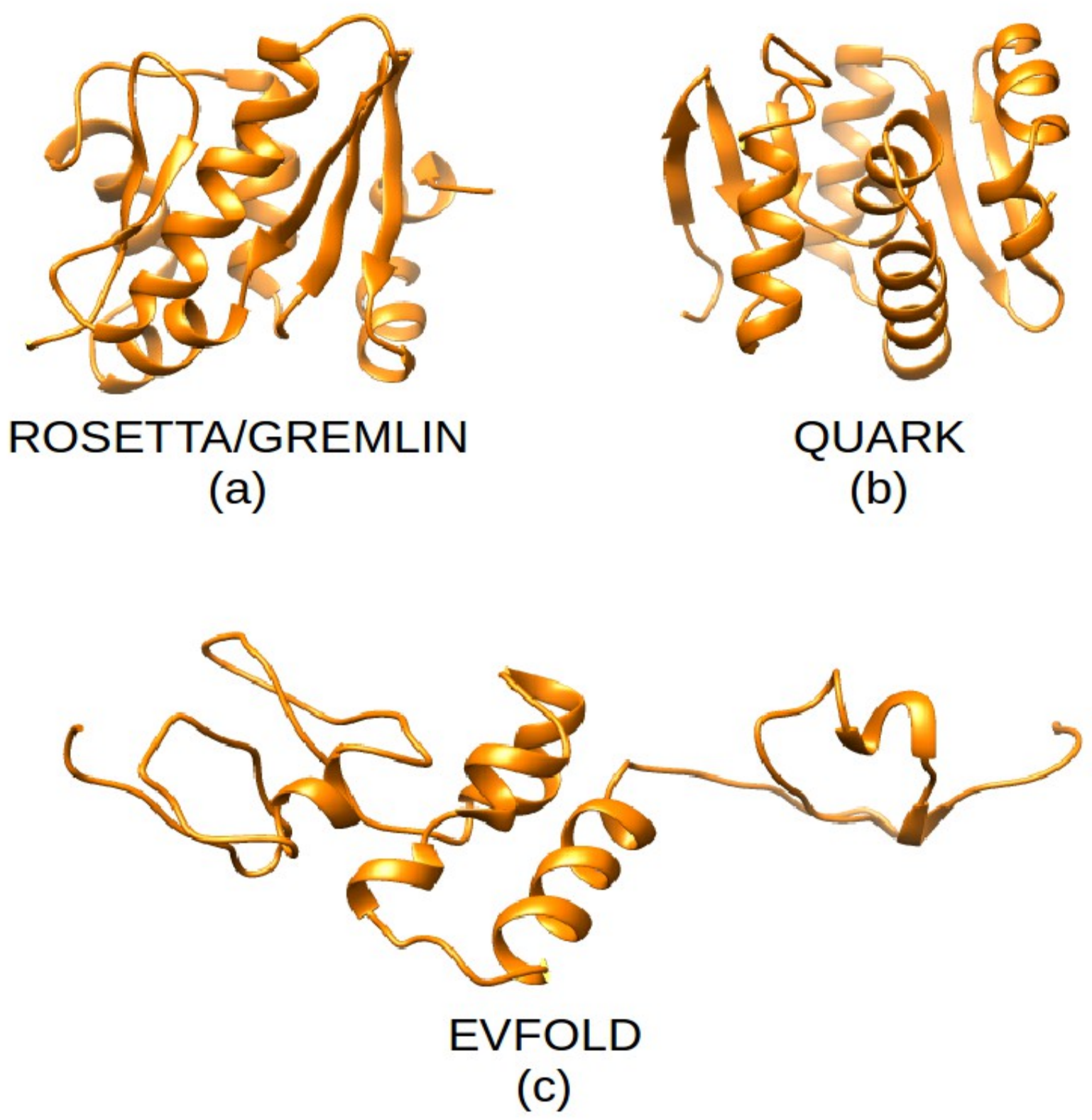

**<u>Supplementary Figure 6</u>**. Models generated using (a) ROSETTA/GREMLIN, (b) QUARK, (c) EVFOLD.

**Supplementary Note 1** (Pages: 33-37)

**Justifications for the Generated Models**

Here we do some simple analysis, based on probability theory, to provide justifications for the 3D models developed.

We assume that the average probability of correctness of a predicted residue-residue contact, used in our model building, is **Pc**=0.4 (from the performance tables in the publications of RAPTORX & CCMPRED – see main text). A predicted contact was chosen for analysis if it satisfied (in the 3D model) αC-αC distance (**d**(αC-αC)) upper bound **Du**. Where, **Du** = 8 Å (upper bound typically used by contact residue predictors, such as RAPTORX) + **t** (tolerance). **t** =4 Å.

Placement of SSES(3):3H in "Modeling Motif-1"

For our analysis we approximate an alpha helix as a line segment, coinciding with the helical axis (depicted in the Figure below).

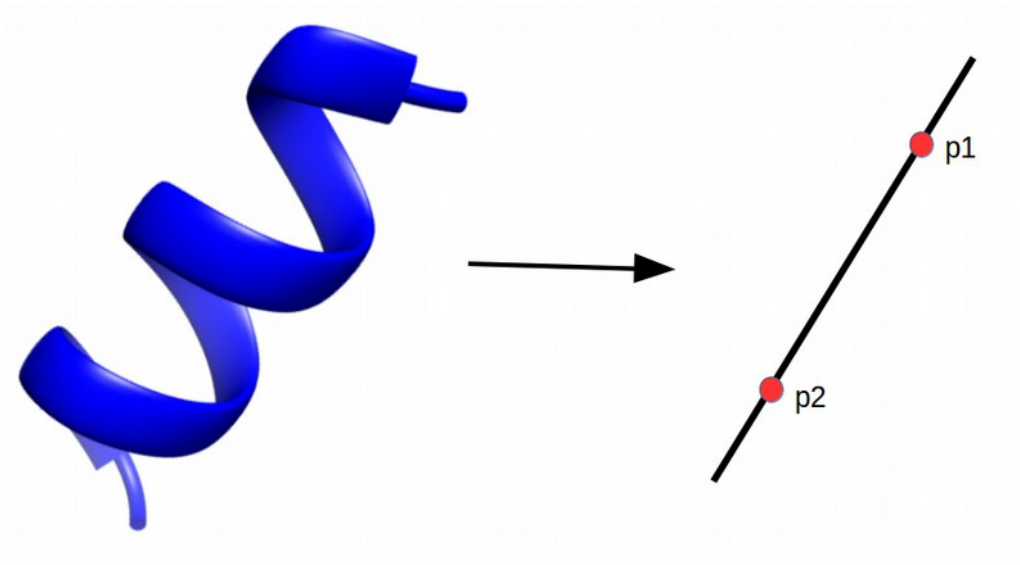

From the theory of spatial geometry, if two distinct points on a line segment are fixed in space then the spatial orientation of the line is also fixed. Let **Pa** be probability that there is at-least one point on the first half of a line segment **L** that is fixed. Let **Pb** be the probability that at least one point on the second half of **L** is fixed. Hence the probability that **L** is fixed in space is **P**=**Pa*****Pb**.

(*continued next page ...*)

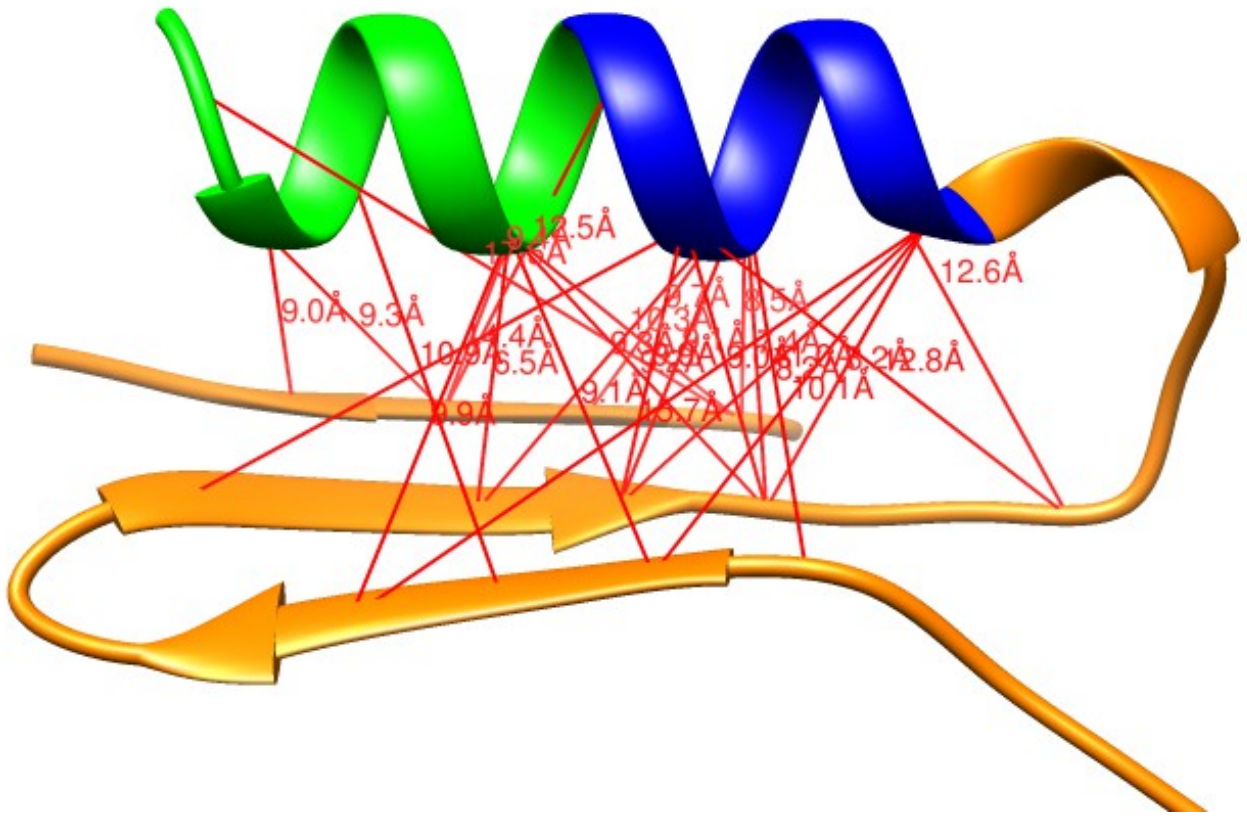

Now consider the Figure above, which shows the placement of 3H, *wrt* beta-sheet group (formed by 1E, 2E), in the final Motif-1 model. Also shown are the predicted residue-residue contacts (via red lines). Since, here we approximate helix as a line segment, **Pa** is the probability that at-least one predicted contact, between the sheet group and the first half of the helix (shown in blue), is true and **Pb** is the probability that at-least one predicted contact, between the sheet group and 2nd half of the helix (shown in green), is true. Here, **Pa**=1-(1-**Pc**)^11=1-(1-0.6)^11=0.99, as we found eleven predicted contacts (with **d**(αC-αC) <= **Du**) in the first half of the helix. **Pb**=1-(1-**Pc**)^8=1-(0.6)^8=0.98, as we found eleven predicted contacts (with **d**(αC-αC) <= **Du**) in the second half of the helix. Hence, here, probability **P** that the helix (approximated as a line segment) is fixed or rigidly-coupled with respect to the sheet group is **Pa*****Pb**=0.99*0.98=0.98 (reasonably high).

<u>Placement of SSES(7):7H in “Modeling Motif-2”</u>

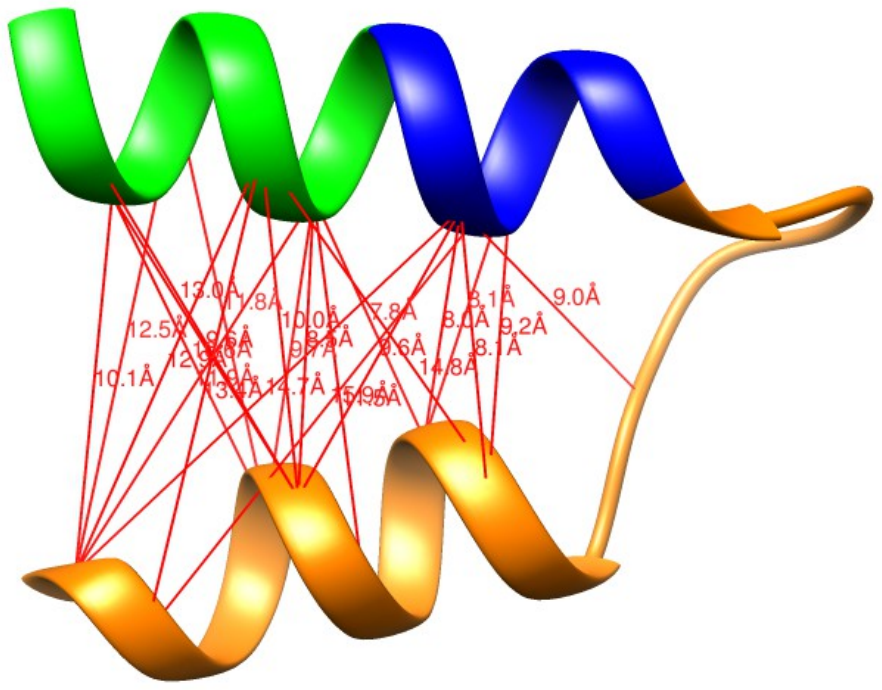

Figure above shows placement of SSES(7):7H (helix at top) *wrt* SSES(6):6H in the final Motif-2 model, along with the predicted residue-residue contacts (via red lines). Using the same kind of analysis as in the previous section, **Pa** (that at-least one contact on first half of helix 7H, shown in blue in Figure above, is true) = (1-(1-**Pc**)^6), as there are six contacts (with **d**(αC-αC) <= **Du**) in the blue region. **Pb** (that at-least one contact on 2nd half of helix 7H, shown in green in Figure above, is true)=(1-(1-**Pc**)^),

as there are nine contacts (with **d**(αC-αC) <= **Du**) in the green region. Hence, **P** (that 7H is fixed *wrt* 6H)= **Pa*****Pb**= (1-(1-**Pc**)^6)*(1-(1-**Pc**)^9)=(1-0.6^9)*(1-0.6^6)=0.95*0.99=0.94 (reasonably high).

<u>Placement of SSES(10):10H in “Modeling Motif-3”</u>

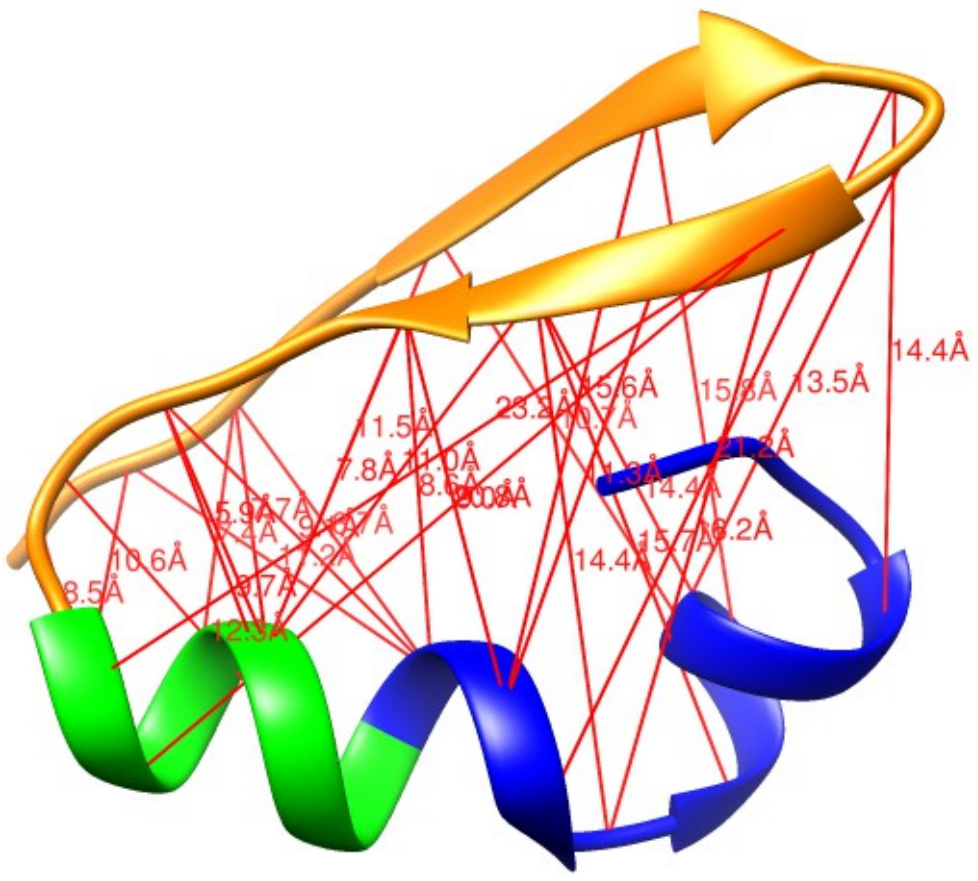

The Figure above shows placement of helix SSES(10):10H *wrt* previous sheet group (formed by 8E, 9E) in the developed model, along with the predicted residue-residue contacts (via red lines). Using the same kind of analysis, as in the previous section, **Pa** (that at-least one contact on first half of helix H: shown in green) = (1-(1-Pc)^7), as there are seven contacts (with **d**(αC-αC) <= **Du**) in the green region. **Pb** (that at-least one contact on 2nd half of helix H: res:-X-Y; shown in blue)=(1-(1-Pc)^7), as there are seven contacts (with **d**(αC-αC) <= **Du**) in the blue region. Hence **P** (that 10H is fixed *wrt* the previous sheet group)=**Pa*****Pb** =(1-(1-**Pc**)^7)*(1-(1-**Pc**)^7)= (1-0.6^7)*(1-0.6^7)=0.94 (reasonably high).

(*continued next page ...*)

Placement of Motif-2 *wrt* Motif-1 in “Modeling Motif-12”

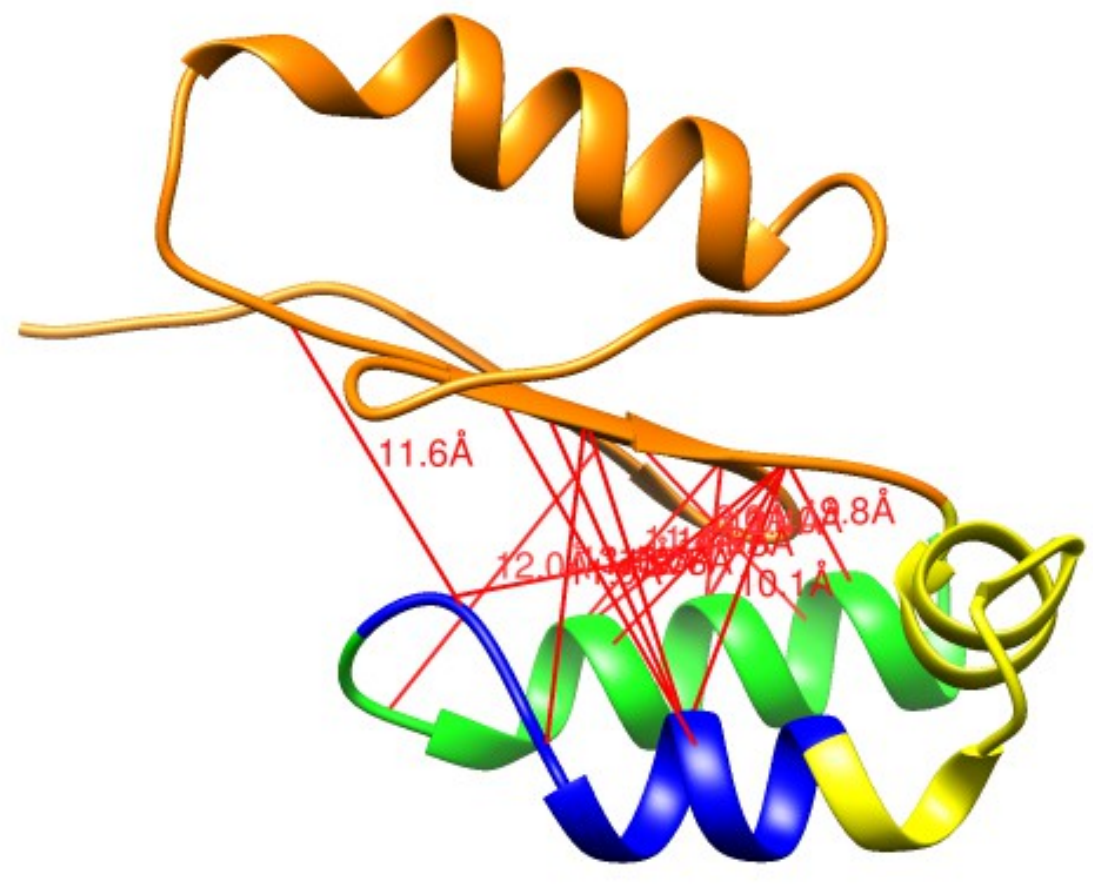

The Figure above shows Moitf-12. The orange coloured part is Motif-1 and rest is Motif-2 along with the joint reside 119. Also shown are predicted contacts between the green/blue region of Motif-2 and the sheet group of Motif-1. Here we estimate the probability that relative position of Motif-2 *wr*t Motif-1 is correct, as determined by contact map predictions. We assume that a residue pair contact holds the residues (in the pair) rigidly in space, with respect to each other. From the theory of spatial geometry, we know a rigid body is rigidly held against another rigid body, if there are three non-collinear points in one rigid body which are held rigidly against three corresponding non-collinear points in the other rigid body. We assume one of the three non-collinear points, needed to show rigidity, is the joint residue between Motif-1 and Motif-2. Let the second non-collinear point be from one of the predicted contacts in the blue region (show in above Figure) of Motif-2. Let the third non-collinear point be from one of the predicted contacts in the green region (shown in above Figure) of Motif-2. The blue and green regions were chosen such that the three non-collinear point, needed to show rigidity of Motif-2 *wrt* Motif-1 would indeed be non-collinear. Let **Pa** be probability that one of the contacts shown in blue portion of Motif-2 is correct. Let **Pb** be probability that one of the contacts shown in green portion of Motif-2 is correct. Then **P**=**Pa***Pb** is the desired probability that Motif-2 is held rigidly against Motif-2, as in the final Motif-12 model. **Pa**=1-{(1-**Pc1**)*(1-**Pc2**)*(1-**Pc3**)*...}, where **Pc*i*** is the probability that a predicted contact shown in the blue portion is correct. **Pc*i***=**Pc**=0.4, if **d**(αC-Cα), distance corresponding to the contact residue pair, <= **Du**. If **d**(Cb-Cb)>**Du**, then we scale down **Pci** as:- **Pci**=**Pc***(**Du**/**d**(Cb-Cb))^2. This yields **Pa**=1-(1-0.4)*(1-0.4)*(1-0.4)*{1-0.4*(12/14.9)^2}*{1-0.4*(12/12.8)^2}*{1-0.4*(12/13.5)^2}*{1-0.4*(12/13.1)^2}=1-0.6*0.6*0.6*0.74*0.65*0.68*0.66=0.95 (using the seven predicted contacts shown in the blue region of the above Figure). Likewise, **Pb**=1-(0.6)^6*{1-0.4*(12/14.6)^2}*{1-0.4*(12/12.4)^2}=1-(0.6)^6*0.73*0.63=0.98 (using the eight predicted contacts shown in the green region of the above Figure). Finally **P**=**Pa***Pb**=0.95*0.98=0.93, the “reasonably high” probability that Motif-2 is indeed held rigidly against Motif-2 as in the Motif-12 model.

Placement of Motif-3 *wrt* Motif-12 in “Modeling Motif-123”

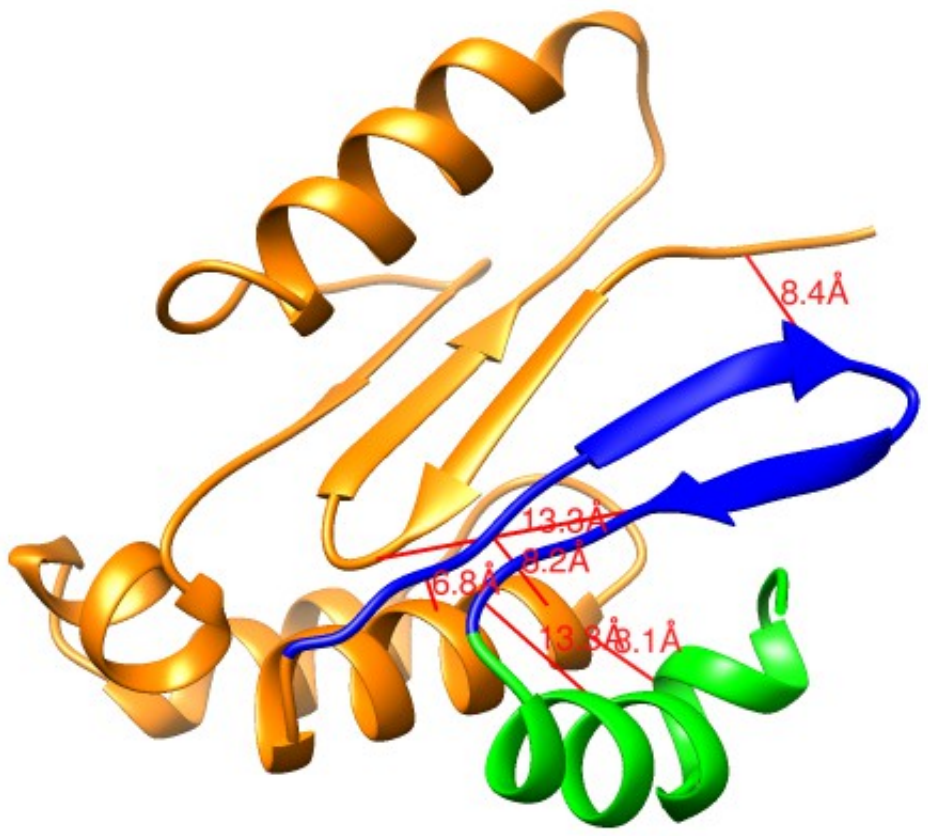

The above Figure shows the placement of Motif-3 (green and blue regions) *wrt* Motif-12 (orange region). The green and blue regions correspond to Motif-3 region breakdown, along the same lines as in previous section. Also shown in the Figure are the predicted contacts (via red lines) between the blue/green region and neighboring SSEs on Motif-12. Using exactly the same kind of steps, as in previous section, we computed the probability that the placement of Motif-3 *wrt* Motif-12 in “Modeling Motif-123” is correct is 0.51. This probability is not high enough, and hence we sampled additional conformations of Motif-3, as described in the main text.

(*End of Supplementary Note 1*)